\UseRawInputEncoding
\documentclass[a4paper,10pt,oneside]{article}
\usepackage[utf8]{inputenc}
\usepackage{icad2024,amsmath,epsfig,times,url,hyperref}
\usepackage[T1]{fontenc}

\usepackage{booktabs}
\usepackage{graphicx}

\title{Flowers Revisited: A preliminary replication of Flowers et al. 1997}

\name{Kajetan Enge $^{1,2,^*}$, Liam Fabry $^1$, Robert Höldrich $^{2,+}$}
    \address{$^1$ St. Pölten University of Applied Sciences, St. Pölten, Austria \\
    $^2$ IEM, University of Music and Performing Arts Graz, Graz, Austria  \\\\
    {\tt $^{*}$kajetan.enge@fhstp.ac.at} \\ 
    {\tt $^{+}$robert.hoeldrich@kug.ac.at}}

\begin{document}
\ninept
\maketitle
\begin{sloppy}
\begin{abstract}
In 1997, Flowers, Buhman, and Turnage published a paper titled ``Cross-Modal Equivalence of Visual and Auditory Scatterplots for Exploring Bivariate Data Samples.'' This paper examined our capacity to assess the relationship between two data variables when presented through visual or auditory scatterplots. Twenty-seven years later, we have replicated the first part of this influential study and present the preliminary findings of our replication, initially involving 21 participants. In addition to purely auditory and visual scatterplots, we introduced audiovisual scatterplots as a third condition in our experiment. Our initial findings mirror those of Flowers et al.'s original research. With this extended abstract, we also aim to spark a discussion about the significance of replication studies for our research community in general.
\end{abstract}

\section{Introduction and Related Work}
\label{sec:intro}

Twenty-seven years after Flowers, Buhman, and Turnage published their seminal paper titled "Cross-Modal Equivalence of Visual and Auditory Scatterplots for Exploring Bivariate Data Samples,"~\cite{flowers_1997_CrossModalEquivalenceVisuala} we present a preliminary replication of their study. By now, the original research has been cited more than 150 times, which makes it one of the most cited sonification papers ever (see Figure 2.b in~\cite{vogt_2023_reflecting} and our supplemental material). In their work, Flowers et al. did study perceptual differences and equivalences of visual and auditory data displays, employing visual and auditory scatterplots. They presented two experiments, one on the human ability to estimate the magnitude of correlations and one on the influence of outliers on such estimations. Their results showed no significant dependency on the used modality, while the variability of estimations was higher for auditory than for visual representations. Also, single outliers had no significant influence on their results. 
We replicated the first part of their study (without outliers) and extended it by an additional condition: audiovisual scatterplots. With the audiovisual condition, participants were able to both see and hear the scatterplots at the same time. 
We consider our replication to be preliminary, as there are small but distinct differences between the original study and our replication, such as differently designed datasets presented to the participants and different playback speeds of the auditory scatterplots. 

%\section{Related work}

% \begin{figure}
%     \centering
%     \includegraphics[width=1\linewidth]{pics/citations.pdf}
%     \caption{Number of citations of Flowers et al.~\cite{flowers_1997_CrossModalEquivalenceVisuala} over the years. While the paper was cited mostly in the early 2000s, it is still a widely cited paper within the sonification community. Data collected from Google Scholar on Feb. 8th 2024}
%     \label{fig:enter-label}
% \end{figure}

While more than 150 papers did cite the study by Flowers et al., one field that has taken relevant inspiration from their work is the field of accessibility. Several papers in the context of accessibility research refer to~\cite{flowers_1997_CrossModalEquivalenceVisuala}, out of which some of the more recent ones are \cite{elavsky_2022_HowAccessibleMy, ferres_2013_EvaluatingToolImprovinga, jiang_2023_UnderstandingStrategiesChallenges, sharif_2021_UnderstandingScreenReaderUsersb,holloway_2022_InfosonicsAccessibleInfographicsa}. The original motivation of Flowers and his colleagues, with the 1997 paper and with others, was to study the needs of assistive technology. In a paper from 2005 where they reflect on their previous research, they nevertheless state that the ``lack of sufficient numbers of visually impaired and blind participants precluded research specifically directed at that population.''~\cite{flowers_2005_DesktopDataSonification}.

With this extended abstract, we contribute results from a preliminary reproduction study of Flowers et al.~\cite{flowers_1997_CrossModalEquivalenceVisuala}, generally mirroring their results with minor deviations only. Our results suggest that Flowers et al. described a phenomenon that seems robust against small differences in study design and, therefore, also appears under modified conditions. Furthermore, we wish to initiate a discussion on the necessity for replication studies within ICAD in general, as well as whether the community is interested in a full-scale replication of the analysis presented by Flowers, Buhman, and Turnage in 1997.

\section{Method}

In this section, we will present our methodology and study design. Overall, Flowers et al. provided all the relevant information that we needed to replicate their study. A few details could not be replicated with full confidence, such as their exact sound design and the axis-scaling of their visual scatterplot (if it was a square or a rectangle they presented). In such cases, we designed our experiment to the best of our knowledge. To help the reader compare to the original study, we use the same structure as Flowers et al.~\cite{flowers_1997_CrossModalEquivalenceVisuala}.

\begin{figure}
    \centering
    \includegraphics[width=1\linewidth]{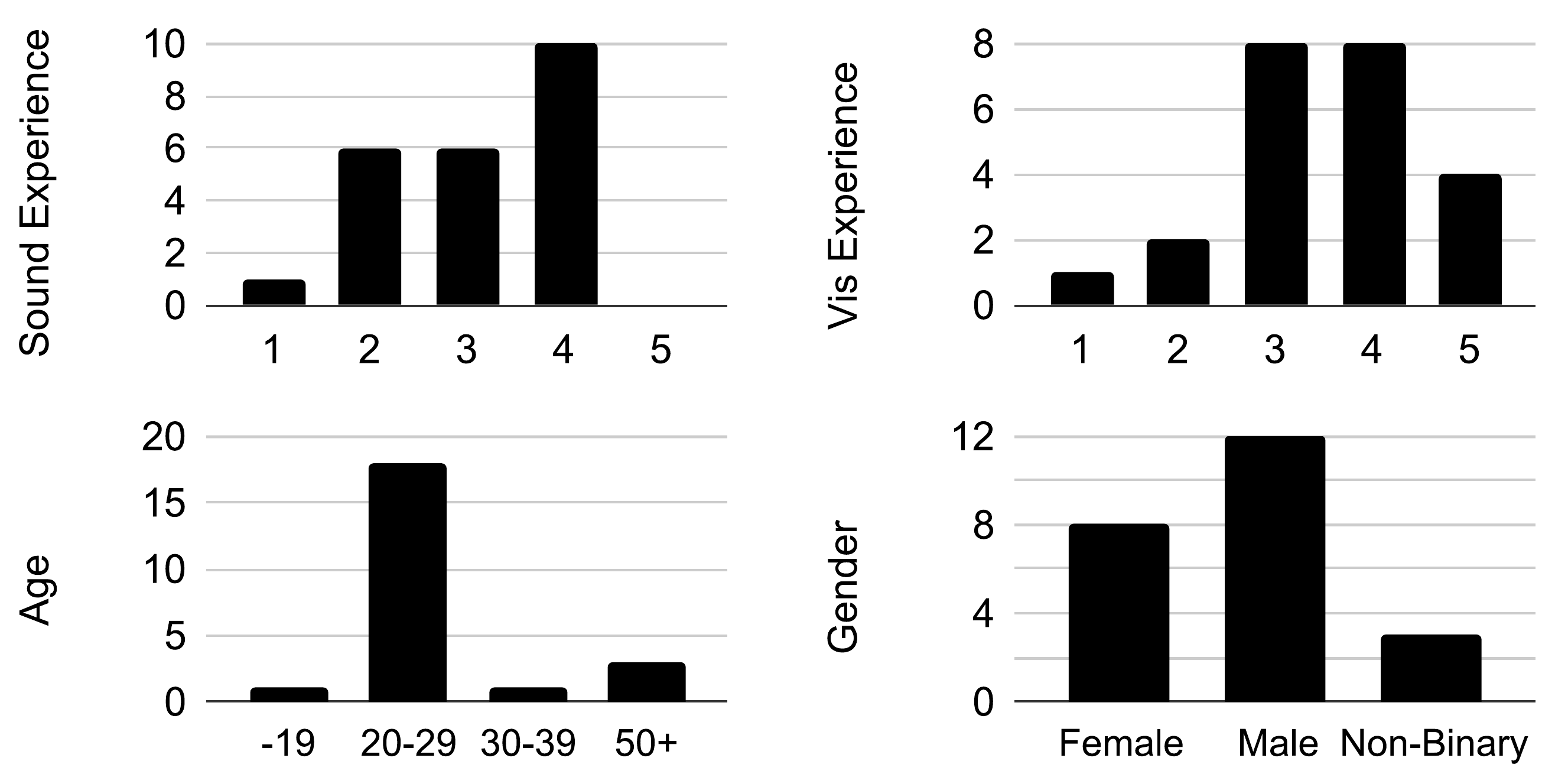}
    \caption{Our participants' self-reported prior experience with sound and visualization, as well as their age and gender. For this preliminary reproduction, we did convenience sampling to recruit participants.}
    \label{fig: stats}
\end{figure}

\vspace{0.2 cm} \noindent
\textbf{Participants:} 23 volunteers took part in this study. \autoref{fig: stats} displays the distribution of their self-reported prior experience with visualization and with music and sound, as well as age and gender. While Flowers et al. had recruited students from an advanced undergraduate psychology class, we recruited colleagues, friends, and family, arguably a sample with different diversity. While Flowers et al. performed a between-subject experiment, we performed a within-subject experiment due to our restricted resources regarding both time and the number of available participants.

\vspace{0.2 cm} \noindent
\textbf{Stimulus material:} 
We used NumPy in a Jupyter Notebook Environment to generate the data samples. Utilizing the function \texttt{numpy.random.multivariate\_normal()}, we sampled 50 numbers from a Gaussian distribution with a mean of 50 and a standard deviation of 10. 

Random sampling was done initially with one seed and ``stretched out'' to generate different magnitudes of positive and negative correlations. Our supplemental material holds a video with a brief animation through the different datasets, visualizing the described ``stretching.'' The resulting datasets hold very similar correlation magnitudes to those in~\cite{flowers_1997_CrossModalEquivalenceVisuala}.

In the original study, the authors computed the datasets for different magnitudes of correlation differently, which we only realized after the experiment. This difference, even if not necessarily influential on our participants' visual or auditory impression, is the first reason for considering our replication preliminary. During the experiment, participants were presented with the stimuli in individual and randomized order, making it impossible for them to draw conclusions from previously seen or heard stimuli. Nevertheless, a full-scale replication should fix this difference and compute the test data like Flowers et al.~\cite{flowers_1997_CrossModalEquivalenceVisuala}.

\vspace{0.2 cm} \noindent
\textbf{Construction of visual scatterplots:} To render the visual scatterplots, we used the \textit{matplotlib} library. All scatterplots were presented on 70x70 unit axes with markers every ten units. We rendered the coordinate system in a square shape, hence without visual distortion between the horizontal and the vertical axes. While we cannot be fully confident that the original study also used square-shaped scaling, we consider non-distorted scaling to be the most intuitive design.

\vspace{0.2 cm} \noindent
\textbf{Construction of auditory scatterplots:} We used SuperCollider and a pre-built plug synthesizer (Karplus-Strong) to generate the auditory scatterplots. The playback speed was set such that the onset times of the 50 data points would fall into a time frame of 3 seconds. Each individual sound had a duration of 0.1 seconds. The pitch mapping was dependent on a point's vertical position, with the lowest value mapped to one octave below the middle C and the highest value mapped to two octaves above the middle C (as in \cite{flowers_1997_CrossModalEquivalenceVisuala}). The auditory scatterplots were rendered to a WAV file and were played back to the participants using Beyerdynamic DT 770 headphones. The scatterplot was played back twice for each audio file with 2 seconds of pause, allowing the participants to listen to one stimulus twice before answering.
The playback time of three seconds (five seconds in~\cite{flowers_1997_CrossModalEquivalenceVisuala}) is the second reason we consider this replication as preliminary. We used a faster playback speed due to a misunderstanding when reading the original study. On the one hand, the different playback speeds hinder direct comparison, but the similarity of our results suggests a robust phenomenon independent of such minor differences in design. 

\vspace{0.2 cm} \noindent
\textbf{Construction of audiovisual scatterplots:} To generate audiovisual scatterplots, we combined the stimuli for visual and auditory scatterplots into a video file. The audio was played back twice in the video, with a break of two seconds in between. During the playback, the visual scatterplot was static, without animation of any kind.

\vspace{0.2 cm} \noindent
\textbf{Procedure:} After filling in a survey about the participant's age, gender, and prior experiences with visualization and sound, each participant was briefly introduced to the concept of correlation and did a training session. During the training, we presented all three types of stimuli with different Pearson correlation values between 1 and -1 to familiarize them with the range of possibilities. During the experiment, participants were presented with the 24 datasets in randomized order. While the stimuli were grouped with respect to the presented modality, different participants started with different modalities to account for potential learning effects. In the original study, the experiments were performed in groups of size three to 16 in a classroom utilizing an overhead projection and desktop computer loudspeakers. We tested every participant individually, utilizing a laptop screen and headphones. The laptop displayed the scatterplots and a user interface. The interface was a $100 mm$ long slider ranging from``Perfect Negative'' to ``Perfect Positive'' with ``Zero'' in between. Flowers et al. used the same visual representation but had a range of $106 mm$ instead. To present our results, we normalized mean and standard deviation values to the $106 mm$ scale. This allows for the direct comparison of the values reported by Flowers et al. while ignoring that our slider was $100 mm$ instead of $106 mm$ wide.

\section{Results}

In this section, we report on the most relevant metrics replicated from the first part of the original study \cite{flowers_1997_CrossModalEquivalenceVisuala}. To help the reader with the comparison to the original study, we copy their paragraph titles.

\vspace{0.2 cm} \noindent
\textbf{Correlations between actual Pearson's correlation and individual participant ratings:} While 23 participants took part in the study, the data of only 21 are considered in our analysis. Two participants had misunderstood the concept of correlation, which only became clear after the experiment. For the remaining participants, the mean correlation between their visual judgments and the true correlation was $r = 0.95$ with a range between $r = 0.84$ and $r = 0.99$. The mean correlation between their auditory judgments and the true correlation was $r = 0.85$ with a range between $r = 0.42$ and $r = 0.95$, and the mean correlation between their audiovisual judgments and the true correlation was $r = 0.95$ with a range between $r = 0.85$ and $r = 0.99$. Hence, we observe high correlations between the participant's estimations and the true values for all three conditions. This mirrors the original results presented in ~\cite{flowers_1997_CrossModalEquivalenceVisuala}, where participants with very low correlations had been excluded from the analysis due to ``clerical errors when recording their [the participants'] judgments.'' 

\begin{figure}
    \centering
    \includegraphics[width=1\linewidth]{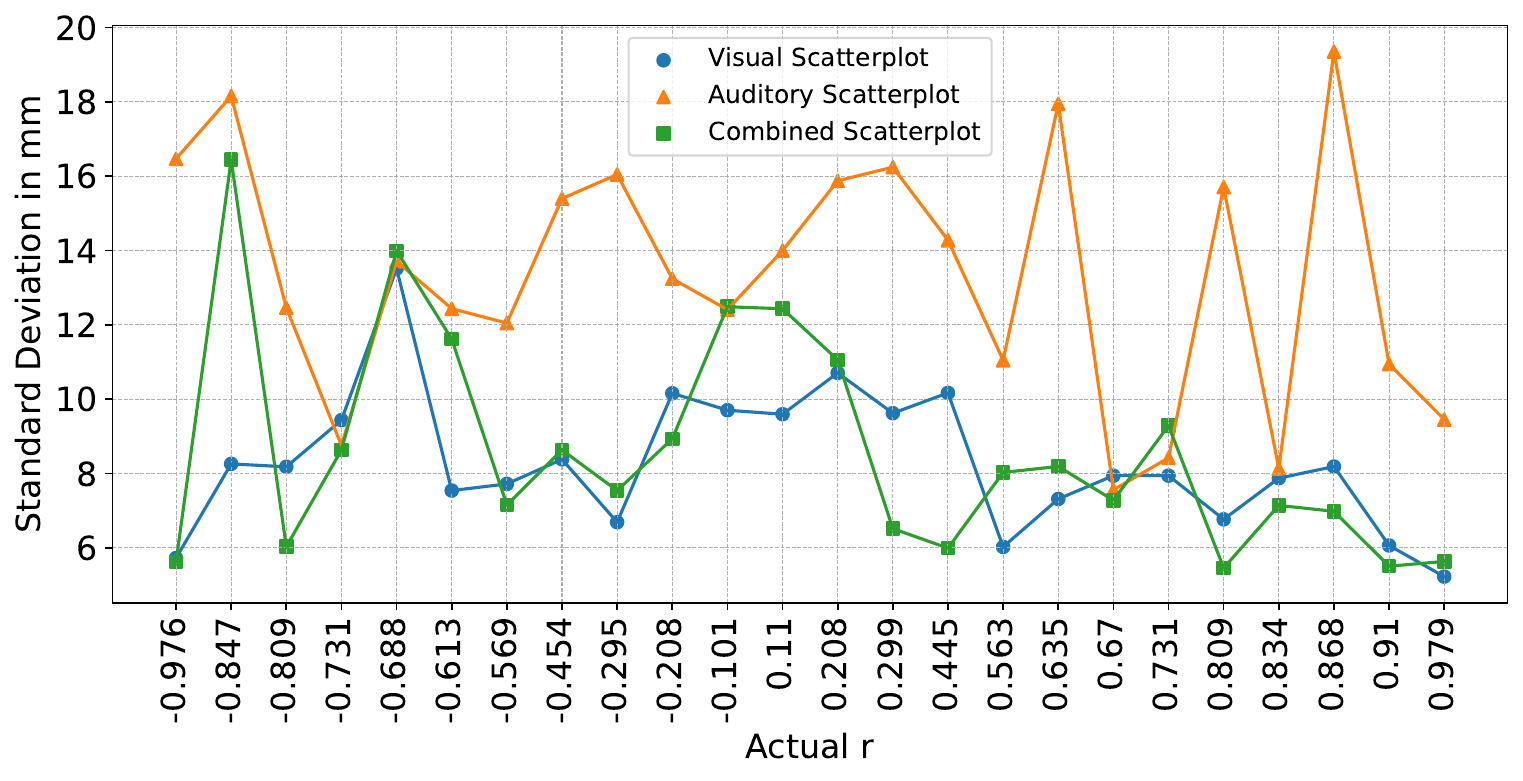}
    \caption{Compare this figure to figure 2 from \cite{flowers_1997_CrossModalEquivalenceVisuala}. The auditory estimations had the largest variability, resulting in a mean standard deviation of $13.34 mm$ (normalized to the $106 mm$ scale used by Flowers et al., who found a mean standard deviation of $10.84 mm$).}
    \label{fig: STDs}
\end{figure}

\vspace{0.2 cm} \noindent
\textbf{Variability of correlation estimates among participants for each modality:} Concerning the variability of the estimates, we found statistically significant differences between the modalities. Note that a Shapiro–Wilk test did not classify each of the response vectors as following a normal distribution, which is why we decided to check all histograms (see supplemental material) visually and decided to run student t-tests as well as their non-parametric alternative: Wilcoxon signed rank tests. 
\autoref{fig: STDs} displays the standard deviations across all 24 levels of correlation. The mean standard deviation is $8.3 mm$ for the visual, $13.3 mm$ for the auditory, and $8.6 mm$ for the audiovisual scatterplot, normalized to the $106 mm$ value range used in~\cite{flowers_1997_CrossModalEquivalenceVisuala}.
Comparing the mean standard deviations between visual and auditory stimuli resulted in a statistically significant difference ($t(24) = 6.4$, $p < 0.001$ \& $w(24) = 8$, $p < 0.001$). Comparing visual and audiovisual stimuli did not reveal statistically significant differences ($t(24) = 4.5$, $p = 0.65$ \& $w(24) = 145$, $p = 0.9$), but comparing auditory and audiovisual stimuli did reveal significant differences ($t(24) = 5.17$, $p < 0.001$ \& $w(24) = 11$, $p < 0.001$). All results are compared to a Bonferroni-corrected $\alpha$-level, necessary due to multiple comparisons. Regarding the comparison between visual and auditory scatterplots, our results mirror the ones presented in the original study. Furthermore, our results suggest that, for audiovisual analysis, the visual modality was the dominant one, resulting in higher variability only when no visualization was available.

\begin{figure}
    \centering
    \includegraphics[width=1\linewidth]{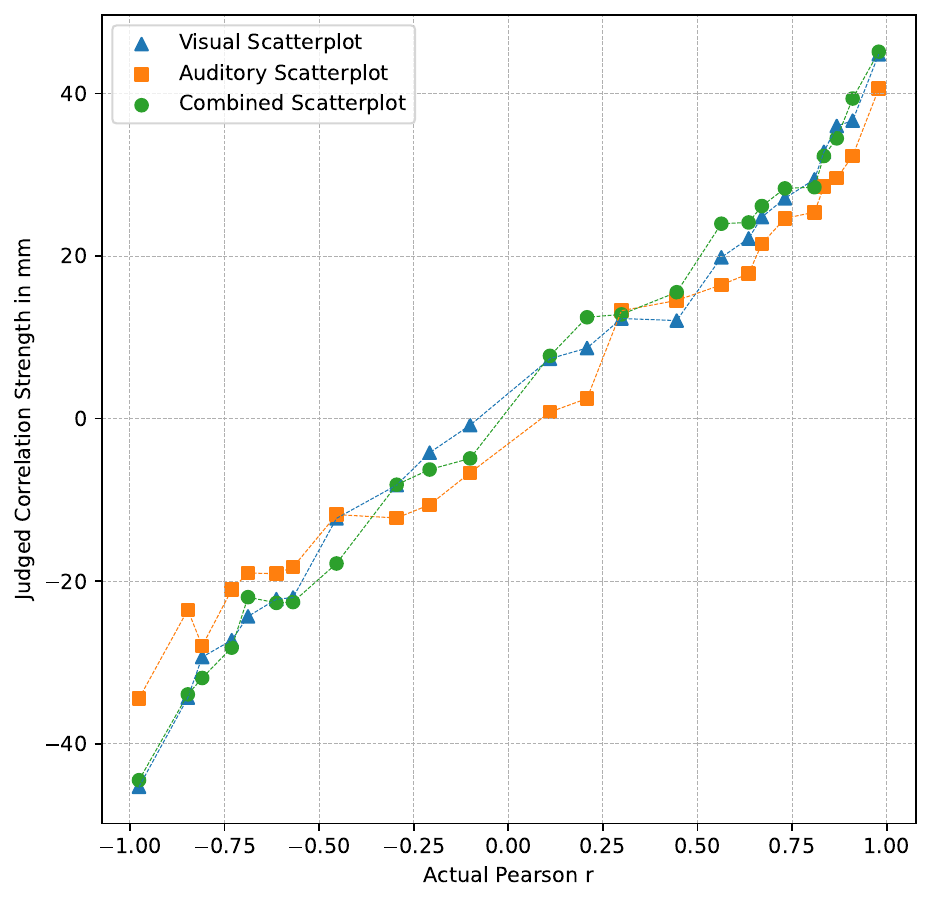}
    \caption{Compare this figure to figure 3 from \cite{flowers_1997_CrossModalEquivalenceVisuala}. For improved readability, we added dashed lines between the data points. The data appears more linear than the data collected in the original study. Nevertheless, the modalities perform similarly to each other, as they did in~\cite{flowers_1997_CrossModalEquivalenceVisuala}.}
    \label{fig: MEANS}
\end{figure}

\vspace{0.2 cm} \noindent
\textbf{Cross-modal equivalence inferred from group data:} \autoref{fig: MEANS} shows the mean correlation estimates plotted against the true correlation values presented to the participants. The figure shows that our participants estimated the level of correlation very similarly regardless of the condition, which, again, mirrors the original study results. Both student t-tests as well as a Wilcoxon signed rank tests comparing all three different modalities did not show significant differences between them (visual vs. auditory:  $t(24) = 0.12$, $p = 0.9$ \& $w(24) = 108$,  $p = 0.24$ | visual vs. audiovisual:  $t(24) = 0.02$, $p = 0.98$ \& $w(24) = 134$,  $p = 0.66$ | auditory vs. audiovisual: $t(24) = 0.15$, $p = 0.88$ \& $w(24) = 114$,  $p = 0.32$).

\section{Discussion}

When sonification researchers refer to \cite{flowers_1997_CrossModalEquivalenceVisuala}, they often argue that the human ability to judge correlations is similar for both visual and auditory representations. Our replication supports such statements and suggests the phenomenon is a robust one. Slight differences in playback time, dataset generation, and sound design did not drastically change our results. Nevertheless, we also want to discuss one notable difference between Flowers et al.'s results and ours: the shape of the psychophysical function.

Compared to \cite{flowers_1997_CrossModalEquivalenceVisuala}, our data suggests a different shape of the psychophysical function between participants' estimations and true correlation values. \autoref{fig: MEANS} shows a more linear relationship between true and estimated correlations than the ones presented in the original study. This phenomenon is related to the ``anchoring effect'' described in the results section of \cite{flowers_1997_CrossModalEquivalenceVisuala}, where medium correlations (around $r = \pm 0.5$) tend to be underestimated in magnitude. Also, the standard deviation (figure 2 in~\cite{flowers_1997_CrossModalEquivalenceVisuala}) shows increased standard deviations of estimations for medium correlations, both for visual and auditory stimuli. In our replication, we do not clearly observe such systematically increased standard deviations for medium correlations, as can be seen in \autoref{fig: STDs}. Hence, while we also observe a monotonic relationship between true and estimated correlations, its curvature seems to differ. 

%Are both of these related to the SEE that is different between the two experiments? See/cite Lane 1985 who stated that SEE is influencial on estimation.
% the reason might lie in different Standard Errors of Esimate (SEE) for the original study and our replication. IN table 1 of~\cite{} the authors report on the SEE values of their displayed data, ... 
%     \item graphically we see a different curvature and pose the question: why? --> Liams table shows overall SEEs that are more equally distributed than those of flowers et al. I have to check this is properly calculated, but this could be the reason for the reduced anchoring effect. If the SEE is lower in the middle part, then it's influence on the perceived relatedness is lower --> hence, the true r is more influencal to the estimation and this will result in an overall more linear relationship between true r and estimated r. 

With respect to the standard deviations of the audiovisual stimuli, we observe another interesting phenomenon. When participants have both the visual and the auditory representation available to estimate the magnitude of a correlation, it seems the visual representation is the dominant one. While the standard deviations for the visual and the audiovisual condition do not differ significantly, the standard deviations for the purely auditory condition are larger. As mentioned already by Flowers et al., the larger variability for auditory stimuli is not surprising and might be explained by the potentially lower auditory literacy of the participants. 

We think of the data presented in this extended abstract as preliminary. In the case of a full-scale replication of the study by Flowers et al.~\cite{flowers_1997_CrossModalEquivalenceVisuala}, we suggest additional analysis methods and perspectives. We argue that model selection methodologies should be used, also employing metrics such as the Akaike Information Criterion (AIK) to account for overfitting. An identification of the best-fitting model to predict the psychophysical function between estimates and true correlations could provide the community with valuable insights for the future development of a sonification design theory. Also, an additional analysis of potential differences between the judgment of positive and negative correlations could be a valuable contribution, as asymmetric results would be plausible due to different masking effects for rising and falling sound sequences.

With this preliminary replication of~\cite{flowers_1997_CrossModalEquivalenceVisuala}, we also intend to revitalize the discussion about the replicability of the studies in our community. Other fields actively discuss the so-called replication crisis, often referring to an article presented by Ioannidis in 2005~\cite{ioannidis_2005_WhyMostPublished}. In the article, Ioannidis argues that ``most published research findings are false'' and presents statistical analysis supporting that claim. In our community, we identified two other articles mentioning the replication crisis in relation to the development of the field of sonification. In 2013, Degara et al.~\cite{degaraquintela_2013_SONEXEVALUATIONEXCHANGE} presented ``SonEX,'' a community-driven platform meant to facilitate the creation and assessment of standardized tasks, aligning with open science principles and promoting reproducible research.
In 2019, in the article ``Eight Components of a Design Theory of Sonification''~\cite{neesEightComponentsDesign2019}, Nees states that there ``is reason to be concerned about the quality standards of user testing in the current sonification literature. Related domains of study have recently experienced a reckoning of sorts regarding the reproducibility and replicability of their findings.'' 

\vspace{0.2 cm} \noindent
At ICAD 2024, we wish to discuss the following questions with the community: (1) Do we, as ICAD, support the replication of our previous findings in our collective future work, possibly in the form of a long paper? If yes, then (2) how can we, as ICAD, establish a culture of reporting on all necessary details in our papers so that other researchers can replicate our studies?  If yes, then (3) and more specifically related to Flowers et al.~\cite{flowers_1997_CrossModalEquivalenceVisuala}, do we consider auditory/visual/audiovisual scatterplots for correlation estimation relevant to our community and, therefore, a valuable target for a full-scale reproduction study?

Last but not least, while Flowers et al. use the phrase ``perceptual equivalence,'' we would like to argue for using the terminology ``communicative equivalence.'' The goal of every data representation, visual or auditory (or utilizing other senses), is communicating data/information to a human. A study that investigates the communicative equivalence of two data displays in the future should also be qualitative, not only reporting on quantitative metrics such as mean estimations or standard deviations.

\section{Supplemental material}
As supplemental material, we provide (1) the raw data of estimations, (2) a video animation of the ``stretching'' of a randomly sampled dataset to achieve different correlation magnitudes, (3) boxplots showing the distributions of the participants' responses, (4) histograms of the estimations on all 72 different stimuli, (5) a bar chart displaying the number of citations of \cite{flowers_1997_CrossModalEquivalenceVisuala} since 1999, and (6) the HTML experiment framework. The supplemental material can be downloaded here: \url{https://phaidra.fhstp.ac.at/detail/o:5543}

\section{ACKNOWLEDGMENT}
\label{sec:ack}
% We would like to thank Alexander Rind, Michael Iber, and Wolfgang Aigner for their valuable input and all the participants who took part in our experiment. This research was funded in part by the Austrian Science Fund (FWF): P33531-N and the Gesellschaft f\"ur Forschungsf\"orderung Nieder\"osterreich (GFF) SC20-006.

We would like to thank the study participants for their time and effort, as well as Alexander Rind, Michael Iber, and Wolfgang Aigner for their valuable feedback. This research was funded in part by the Austrian Science Fund (FWF): 10.55776/P33531 and the Gesellschaft f\"ur Forschungsf\"orderung Nieder\"osterreich (GFF) SC20-006.

% -------------------------------------------------------------------------
\bibliographystyle{IEEEtran}
\bibliography{refs2024}

\end{sloppy}
\end{document}